\documentclass[prl, superscriptaddress, twocolumn,showpacs, floatfix, nobalancelastpage]{revtex4}
\usepackage{epsfig,amssymb,amsmath}
\usepackage[hypertex,linkcolor=red]{hyperref}

\usepackage{color}
\newcommand{\bra}[1]    {\langle #1|}
\newcommand{\ket}[1]    {| #1 \rangle}
\newcommand{\mat}[2]{
\left(\begin{array}{#1}
#2
\end{array}
\right)}

\renewcommand{\t}[1]{\textrm{#1}}

\def\labell#1{\label{#1}}
\def\>{\rangle}\def\<{\langle}
\def\togli#1{}

\begin{document}
\title{
  Using entanglement against noise in quantum metrology
  }
\author{Rafal Demkowicz-Dobrza\'nski}
\affiliation{Faculty of Physics, University of Warsaw, ul. Ho\.{z}a
  69, PL-00-681 Warszawa, Poland}
\author{Lorenzo Maccone}
\affiliation{Dip.~Fisica and INFN Sez.~Pavia, Univ.~of Pavia, via Bassi 6,
    I-27100 Pavia, Italy}
\begin{abstract}
  \togli{ In unitary parameter estimation one can achieve a
    Heisenberg-like $1/N$ scaling in precision in two ways: sampling
    the system in parallel with $N$ probes in a joint entangled state,
    or sampling it $N$ times sequentially with a single probe. The
    latter strategy does not employ entanglement, but both sample the
    system the same number of times and achieve the same precision.
    Their yield becomes inequivalent when noise is added and,
    surprisingly, the entangled strategy tolerates noise better
    (i.e.~is more precise) than the entanglement-free one for many
    common noise models, an effect that was never observed before in
    quantum-enhanced protocols.  We also analyze the asymptotic
    benefits of entangling ancillary systems in the estimation
    procedure. It is known that ancillary systems are useless in the
    noiseless scenario. We show that this is also the case for the
    erasure and dephasing models but, surprisingly, not for the
    amplitude-damping noise where ancilla-enhanced strategies permit a
    better precision. We use these results to conjecture a general
    hierarchy for quantum metrology strategies in the presence of
    noise.}  We analyze the role of entanglement among probes and with
  external ancillas in quantum metrology. In the absence of noise, it
  is known that unentangled sequential strategies can achieve the same
  Heisenberg scaling of entangled strategies and that external
  ancillas are useless. This changes in the presence of noise: here we
  prove that entangled strategies can have higher precision than
  unentangled ones and that the addition of passive external ancillas
  can also increase the precision. We analyze some specific noise
  models and use the results to conjecture a general hierarchy for
  quantum metrology strategies in the presence of noise.
\end{abstract}
\pacs{03.65.Ta,06.20.-f,42.50.Lc}
\maketitle

Quantum metrology \cite{science,natphot} describes parameter
estimation techniques that, by sampling a system $N$ times, achieve
precision better than the $1/\sqrt{N}$ scaling of the central limit
theorem of classical strategies. Different schemes can beat such limit
(Fig.~\ref{f:schemes}): (i)~entanglement-free classical schemes where
$N/n$ independent probes sense the system sequentially thus rescaling
the parameter, and hence the error, by $n$ for each probe
\cite{luis,pryde}; (ii)~entangled parallel schemes that employ a
collective entangled state of the $N$ probes that sample the system in
parallel \cite{caves1,caves2,helstrom,holevo}; (iii) passive ancilla
schemes, where the $N$ probes may also be entangled with noiseless
ancillas; (iv)~active ancilla-assisted schemes (comprising all the
previous cases) that also encompass all schemes employing feedback:
adaptive procedures are described as unitary operations acting on the
probes and ancillas between the sensing and the final measurement
\cite{qmetr,mosca}.

\begin{figure}[ht]
\begin{center}
\epsfxsize=.9\hsize\leavevmode\epsffile{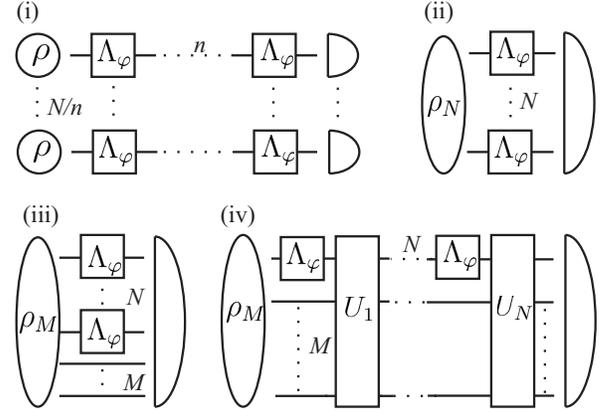}
\end{center}
\vspace{-.5cm}
\caption{ Quantum metrology strategies. The maps $\Lambda_\varphi$
  encode the parameter $\varphi$ to be estimated.  (i)
   sequential scheme: $\Lambda_\varphi$ acts $n$
  times sequentially on $N/n$ input probes $\rho$ (this is an
  entanglement-free classical scheme) ; (ii) entangled
  parallel scheme: an entangled state of $N$ probes $\rho_N$ goes
  through $N$ maps $\Lambda_\varphi$ in parallel; (iii) passive
  ancilla scheme: the $N$ probes are also entangled with $M$ noiseless
  ancillas; (iv) active ancilla-assisted scheme: the action of $N$
  channels $\Lambda_\varphi$ is interspersed with arbitrary unitaries
  $U_i$ representing interactions of the probe with ancillas. [All the other schemes can be derived from (iv) choosing swap or
  identity unitaries $U_i$].
}
\labell{f:schemes}\end{figure}

In the noiseless case, classical single-probe sequential schemes
(i) can attain the same $1/N$ precision as parallel entangled ones
(ii) at the expense of an $N$-times longer sampling time, whereas
passive and active ancilla schemes (iii) and (iv) offer no additional
advantage \cite{qmetr, childs}.  In this paper we analyze the performance of these strategies in the
presence of specific noise models, and use the results to conjecture a
general hierarchy of protocols.
Noise in quantum metrology has been
extensively studied, e.g.~see \cite{chiara,luiz,guta, kitagawa,
  shaji2007, dorner2009,caves1981, xiao1987,
  rubin2007,rafal,kolodynski2010,knysh2010,iran,sahar,matteo1, matteo2}, but the main focus was
  on comparing parallel-entangled with parallel-unentangled strategies \cite{guta,chiara,luiz}  which do not match in the noiseless case.
Single probe states are typically
less sensitive to decoherence and simpler to prepare than
entangled states, so it would seem \cite{luis} that the
sequential strategy should be preferable in the presence of noise.
Out first result is that this is not true:
in presence of noise (here we analyze
dephasing, erasure and damping) entanglement among probes increases the precision over the
sequential strategiy, even thought it fails to do so in the noiseless case,
and we provide a quantitative characterization of this advantage. Our second result is to show that (ii) and (iii) are
in general asymptotically inequivalent, by demonstrating that (iii) is
strictly better than (ii) for amplitude-damping noise. Our third
result is to show that the bounds to parallel-entangled strategies
(ii) and (iii) derived for a large class of noise models \cite{luiz,
  guta} apply asymptotically in $N$ also to the most general
strategies (iv), suggesting that active ancilla-based schemes are not
helpful in increasing the precision in the presence of noise
\footnote{Our claim is stronger than the one in \cite{luiz} that if
  (ii) and (iii) are limited by a $1/\sqrt{N}$ precision scaling, so
  is (iv): we derive an explicit bound for (iv) and show that it
  asymptotically coincides with the bound for (ii) and (iii).}.
Finally, we use our results to
conjecture a general hierarchy of quantum metrology schemes valid in presence of any uncorrelated noise 
\begin{equation}
\label{eq:conjecture}
\begin{array}{cc}
  \t{(i)} = {\t{(ii)}} =  {\t{(iii)}} ={\t{(iv)}} & \quad \t{decoherence free}, \\
  \t{(i)} < {\t{(ii)}}  =  {\t{(iii)}} ={\t{(iv)}} &\quad \t{dephasing, erasure},\\
  \t{(i)} < \t{(ii)} <  {\t{(iii)}}  \overset{?}{=} \t{(iv)}&\quad \t{amplitude-damping}, \\
  \textbf{(i)}\overset{?}{\leqslant}{\textbf{(ii)}} \leqslant {\textbf{(iii)}} \overset{?}{=} {\textbf{(iv)}}  &\quad \textbf{general conjecture}.
\end{array}
\end{equation}
Namely, in general, sequential strategies (i) are worse \footnote{But
  they are equivalent in the noiseless case.} than parallel-entangled
ones (ii), which might in some cases be improved by entangling the
probes with noiseless ancillas (iii), but there is no additional
asymptotic gain from using active ancilla-aided schemes (iv). Question
marks represent our conjectures and the equality symbol ``$=$'' should
be interpreted as asymptotically equivalent, though in the
decoherence-free case as well as in the case of equality between (ii)
and (iii) for erasure and dephasing noise this is a strict
equality for any finite $N$.

Schemes that employ quantum-error correction \cite{kessler2014,
  dur2014, arrad2014} are in general of type (iv), so our claim might be
misinterpreted as saying that error correction schemes are useless.
Instead, what we say is simply that their asymptotic precision can
also be achieved through (possibly unknown) strategies of type
(ii,iii): e.g.~the noise models considered in \cite{kessler2014,
  dur2014, arrad2014} allow for decoupling the decoherence from the
parameter sensing transformation at short evolution times: so, the
bounds derived for (ii,iii) also allow for the possibility of better
than $1/\sqrt{N}$ scaling \cite{barcelona}.

Outline of the paper: we first introduce the quantum Cramer-Rao bound
for the strategies (i-iv), and derive some general bounds for their
quantum Fisher information. We then prove a gap in precision between
(i) and (ii), the equivalence of (ii), (iii), and (iv) in case of
dephasing and erasure noise and finally inequivalence of
(ii) and (iii) for amplitude-damping.

The map $\Lambda_\varphi$ that writes the parameter $\varphi$ on the
state $\rho$ of the probe acts as
\begin{equation}
\label{eq:rhovarphi}
\rho_\varphi = \Lambda_\varphi(\rho) = \sum_k K_k^{\varphi} \rho K_k^{\varphi \dagger},
\end{equation}
with $K_k^{\varphi}$ the Kraus operators. The precision of an
estimation strategy can be gauged through the root mean square error
$\Delta\varphi$ of the measurement of $\varphi$.  It is lower-bounded
by the quantum Cramer-Rao bound
\cite{caves1,caves2,natphot,helstrom,holevo},
$\Delta\varphi\geqslant 1/\sqrt{\nu F(\rho_\varphi)}$, where $\nu$ is the number of times the estimation is repeated, and
$F(\rho)$ is the quantum Fisher information (QFI) of a state $\rho$
\cite{caves1,caves2,natphot}.  This bound  is guaranteed to be
achievable in general only asymptotically for $\nu\to\infty$, but in
case of  noise  models with QFI scaling linearly with the number of probe particles $N$
it is also tight for a single shot setting, $\nu=1$, provided one considers the asymptotics, $N \rightarrow \infty$ \cite{jarzyna2014}.

 The QFIs for the schemes (i-iv) are
defined as
\begin{align}
F^{(\t{i})} & = \max_{\rho, n}F\{[\Lambda^n_\varphi(\rho)]^{\otimes N/n}\} \\
F^{(\t{ii})} & = \max_{\rho_N}F[\Lambda_\varphi^{\otimes N}(\rho_N)] \\
F^{(\t{iii})} & = \max_{\rho_M}F[\Lambda_\varphi^{\otimes N}\otimes \openone^{\otimes M} (\rho_M)] \\
F^{(\t{iv})} & = \max_{\rho_M,\{U_i\}}F[U_N \Lambda_{\varphi} \, \dots U_1 \, \Lambda_{\varphi}(\rho_M)],
\end{align}
where $\rho$ denotes an input state of a single probe and we look for
the optimal sequential-parallel splitting of the $N$ probes in $n$
channels for strategies (i), $\rho_N$ is the global state of $N$
probes in (ii), while $\rho_M$ denotes the global probes-ancilla input
state in (iii) and (iv). In the formula for $F^{\t{(iv)}}$, the $U_i$s
act on all the probes while $\Lambda_\varphi$ without loss of
generality may be assumed to act on the first probe only.  Due to the
convexity of the QFI, the optimal input probes are pure.

The hierarchy conjecture \eqref{eq:conjecture} should be understood in
terms of corresponding inequalities on QFIs: $F^{\t{(ii)}} \leq
F^{\t{(iii)}}$ is obvious as $\t{(ii)}$ is a special case of
$\t{(iii)}$, the inequality may be strict as is the case of the
amplitude-damping discussed below; $F^{\t{(iii)}} \leq F^{\t{(iv)}}$
is also easy to show since taking swap operators $U_i$ in (iv) one can
obtain the action of parallel channels on an entangled input state
(iii). It is less trivial to determine the cases when inequalities
turn to equalities and the corresponding schemes become asymptotically
equivalent.  Finally, the $F^{\t{(i)}} \leq F^{\t{(ii)}}$ inequality
in more challenging to prove in general,
but we show that it holds strictly for dephasing,
erasure or amplitude damping,
proving the advantage of parallel schemes \footnote{This claim may seem in
  contradiction with the one of \cite{shaji,boixo,boixo2} where
  equivalence of the sequential and parallel-entangled strategy is
  proven in for dephasing. The contradiction is only apparent, as no
  optimization over the input state is performed there: only the
  response of the channel is analyzed.}.
We also present general tools to derive bounds for (iv) and show that
they are asymptotically equivalent to known bounds for (ii,iii).
Moreover, since these bounds are saturable for dephasing and erasure using (ii) schemes,
there is no asymptotic advantage of (iv) over the simpler (ii) and
(iii) in these cases.

Calculating QFI explicitly for large $N$ is in general not possible
but bounds to it are known. The most versatile ones employ the
non-uniqueness of the Kraus representation \cite{fujiwara2008, luiz,
  guta}: $\Lambda_\varphi$ is unchanged if one replaces $K^\varphi_k$
with $\tilde{K}_k^\varphi = \sum_l u_{kl}^\varphi K_l^\varphi$, where
$u^\varphi$ is an arbitrary $\varphi$-dependent unitary matrix. This
produces bounds on the the maximal QFI of a transformation
$\Lambda_\varphi$ in terms of minimization over the possible Kraus
representations \cite{fujiwara2008, guta}:
\begin{equation}
  \label{eq:minkraus}
  \max_{\rho}F[\Lambda_\varphi(\rho)] \leq 4 \min_{\{K^{\varphi}_k\}}
  \| \sum_k \dot{K}^{\varphi \dagger}_k \dot{K}^{\varphi}_k \|,
\end{equation}
where $\dot{K}^{\varphi}_k = \frac{\partial K^{\varphi}_k}{\partial
  \varphi}$ and $\| \cdot \|$ is the operator norm.  The above
inequality becomes an equality provided one replaces $\Lambda_\varphi$
with a trivially-extended channel $\Lambda_\varphi \otimes \openone$
which represents the possibility of entangling the probes with an
ancilla \cite{fujiwara2008}. This immediately implies that the bounds
derived for (ii) will also be valid for (iii).

We now recall known bounds for $F^{\t{(ii/iii)}}$ and derive a new bound for $F^{\t{(iv)}}$ using the
minimization of Eq.~\eqref{eq:minkraus}.  Bounds for (ii) and (iii)
are equivalent (as argued above) so we use a combined notation
(ii/iii).  For any Kraus representation $K_k^\varphi$ of a single
channel $\Lambda_\varphi$ one can write a \emph{product} Kraus
representation for channels $\Lambda^{\otimes N}_\varphi$,
$U_1\Lambda_\varphi\,\dots\,U_N \Lambda_\varphi$ corresponding to
schemes (ii/iii), (iv) respectively:
$K^{\varphi (\t{ii/iii})}_{\boldsymbol{k}} =
K^\varphi_{k_N}\otimes \dots \cdot K^\varphi_{k_1}$; $K^{\varphi
  (\t{iv})}_{\boldsymbol{k}} = U_N K^\varphi_{k_N}\cdot \dots \cdot
U_1 K^\varphi_{k_1}$, where $\boldsymbol{k} = \{k_1,\dots, k_N\}$.

For (ii/iii) the minimization \eqref{eq:minkraus} gives a simple bound
expressed in terms of single channel Kraus operators
\cite{fujiwara2008}:
\begin{equation}
\label{eq:boundsii}
F^{\t{(ii/iii)}} \leq 4 \min_{K_k^\varphi} N \|\alpha\| + N(N-1) \|\beta\|^2 \leq  4 \min_{K_k^\varphi, \beta=0} N \|\alpha\|,
\end{equation}
with $\alpha \equiv \sum_k \dot{K}^{\varphi \dagger}_k
\dot{K}^{\varphi}_k$ and $\beta\equiv \sum_{k} \dot{K}^{\varphi
  \dagger}_k {K}^{\varphi}_k$. The last inequality in
\eqref{eq:boundsii} may be used without loss of efficiency for large
$N$ provided there is a Kraus representation for which $\beta=0$ (it
 exists for many noisy maps), which immediately implies linear QFI scaling
with $N$ \cite{fujiwara2008, guta}. The minimization in
Eq.~\eqref{eq:boundsii} can be easily performed using the semi-definite
programming \cite{guta, kolodynski2013}.

The derivation of the general bound for (iv) uses again
\eqref{eq:minkraus} and a product Kraus representation. It gives (see
supplemental material for the details)
\begin{multline}
\label{eq:boundiii}
F^{\t{(iv)}} \leq 4 \min_{K_k^\varphi} N \|\alpha\| + N(N-1) \|\beta\|
(\|\alpha\| + \| \beta\| + 1) \\ \leq 4 \min_{K_k^\varphi, \beta=0} N
\|\alpha\|.
\end{multline}
Importantly, the asymptotic form of the bound is equivalent to
\eqref{eq:boundsii}, the one derived for (ii/iii) if $\beta=0$ is feasible.

It is worth noting, that less powerful but more intuitive methods
based on the concept of minimization over classical or quantum
simulations of the channel \cite{matsumoto2010, guta, kolodynski2013},
originally proposed to derive bounds for (ii/iii), can also be applied to (iv).
\begin{figure}[ht]
\begin{center}
\epsfxsize=.8\hsize\leavevmode\epsffile{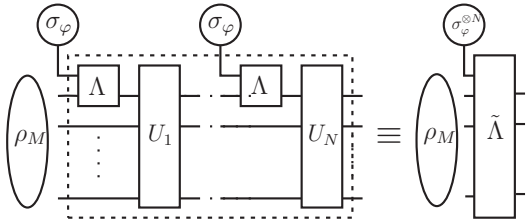}
\end{center}
\vspace{-.5cm}
\caption{Depiction of quantum channel simulation applied to the most
  general adaptive scheme (iv). It shows that, for a given simulation
  $\Lambda, \sigma_\varphi$, the QFI of the scheme is bounded by
  $F^{\t{(iv)}} = F[{\Lambda}(\rho_M \otimes \sigma_\varphi^{\otimes
    N})]\leq N F(\sigma_\varphi)$.}
\labell{f:simulation}\end{figure}
In classical-and-quantum-simulation method \cite{matsumoto2010}
one formally replaces the action of 
$\Lambda_\varphi$ with a parameter-independent map $\Lambda$
and a parameter-dependent ancillary
system $\sigma_\varphi$, so that for any $\rho$:
$\Lambda_\varphi(\rho) = \Lambda(\rho \otimes \sigma_\varphi)$. Since
QFI is nonincreasing under parameter independent maps, $F(\Lambda(\rho \otimes \sigma_\varphi)) \leq
F(\sigma_\varphi)$, 
which for the schemes (ii/iii) implies that
$F[\Lambda_\varphi^{\otimes N}(\rho_N)] \leq F(\Lambda^{\otimes
  N}(\rho_N \otimes \sigma_\varphi^{\otimes N})] \leq N
F(\sigma_\varphi)$ \cite{matsumoto2010, guta, kolodynski2013}.
It has not been noticed before that the same method can be
applied to (iv)
as the scheme can be
rewritten as one black-box quantum operation ${\tilde{\Lambda}}$ fed
with $\sigma_\varphi^{\otimes N}$,
see Fig.~\ref{f:simulation}, resulting in:
\begin{equation}
\label{eq:simulation}
F^{\t{(ii-iv)}} \leq N \min_{\Lambda, \sigma_\varphi} F(\sigma_\varphi).
\end{equation}
This bound often coincides with the asymptotic bound in \eqref{eq:boundsii}, e.g. in the case of
erasure or dephasing \cite{guta, kolodynski2013} (but not amplitude-damping, see \cite{kolodynski2013}).

We now analyze dephasing, erasure and amplitude-damping noise. Let $\ket{0}, \ket{1}$ be
the eigenbasis of the phase encoding unitary $U_\varphi =
\ket{0}\bra{0}+ e^{i \varphi}\ket{1}\bra{1}$.  We assume that the
dephasing is defined with respect to the same basis so the
corresponding Kraus operators read:
\begin{eqnarray}
K_0=\openone\Big({\frac{1+\sqrt{\eta}}2}\Big)^{1/2}\;,\
  K_1=\sigma_z\Big({\frac{1-\sqrt{\eta}}2}\Big)^{1/2}
\labell{cpmap}\;,
\end{eqnarray}
where $\openone=|0\>\<0|+|1\>\<1|$, $\sigma_z=|0\>\<0|-|1\>\<1|$, and
$\sqrt{\eta}$ is the decoherence rate of the off-diagonal terms in the
density matrix. Since both Kraus operators commute with the unitary
$U_\varphi$ we can separate the noise map from the sampling and
consider a total evolution of the form
\begin{eqnarray}
\rho_\varphi= \Lambda_\varphi(\rho)=\sum_k K_k U_\varphi\rho {U_\varphi}^\dag
{K_k}^\dag\;.
\labell{totev}\;
\end{eqnarray}
Instead, for erasure noise the probe is untouched with probability
$\eta$ while with probability $1-\eta$ its state is replaced with one
in a subspace orthogonal to the subspace where the estimation takes
place (again the noise map and $U_\varphi$ commute and the map can be
written in a Kraus form, see supplemental material). The erasure map
is isomorphic to optical loss applied to a state with fixed number of
distinguishable photons with transmission coefficient $\eta$ in both
arms of an interferometer \cite{guta, demkowicz2014}. Finally, Kraus operators for amplitude-damping read
\begin{eqnarray}
\label{eq:krausdamping}
 K_0 = \mat{cc}{1 & 0 \\ 0 & \sqrt{\eta}}\;,\
  K_1=\mat{cc}{ 0 & \sqrt{1-\eta} \\ 0 & 0}
\;,
\end{eqnarray}
where $\eta$ represents the probability of a particle to switch from the excited to the ground state.

We start with calculating  $F^{\t{(i)}}$ to assess the performance of entanglement-free strategies.
 In case of the erasure, since in the
noiseless case the optimal probe state is $\ket{+}=(\ket{0} +
\ket{1})/\sqrt{2}$, while the probability of erasure event does not depend on the state itself,
the optimal input state remains the same, and yields
$F[\Lambda_\varphi(\ket{+}\bra{+})]=\eta$.  For dephasing and amplitude damping the
situation is less obvious but the optimal probe state is again
$\ket{+}$ and the QFI  is
again 
$\eta$ \cite{kolodynski2013}, see supplemental material for a simple proof in case of dephasing.

To calculate $F^{\t{(i)}}$ it remains to optimize the number $n$ of
sequential maps for each probe, see Fig.~\ref{f:schemes}. Using $n$
maps in a sequence increases the overall phase rotation $n$ times at
the cost of increasing the decoherence parameter ${\eta}$ to
${\eta}^n$, whereas considering parallel channels simply adds their
QFIs.  Therefore, $F[\Lambda^n_\varphi(\rho_+)]^{\otimes N/n}\}]= N/n
\cdot n^2 {\eta}^n$.  This is the same formula which would be obtained
for (ii) with input $N00N$ state \cite{dorner2009,shaji,boixo2,mio}.
Treating $1 \leq n \leq
N$ as a continuous parameter \cite{dorner2009}, the optimal value $n=
[\ln(1/{\eta})]^{-1}$, provided $ e^{-1} \leq {\eta} \leq e^{-1/N}$,
which corresponds to \footnote{For completeness we should add that if
  ${\eta}< e^{-1}$ we take $n=1$ which gives $F= N {\eta}$, while for
  ${\eta} > e^{-1/N}$ we take $n=N$ which gives $F= N^2 {\eta}^N$.
  Note that in the asymptotic limit $N\rightarrow \infty$ and
  ${\eta}<1$, we can ignore the case ${\eta} > e^{-1/N}$.}
\begin{equation}
\label{eq:seqprec}
F^{\t{(i)}} = \frac{N}{e \ln(1/{\eta})}.
\end{equation}

For erasure and dephasing,  we use the inequality \eqref{eq:boundsii} to calculate (see supplemental material)
\begin{equation}
\label{eq:entbasedbound}
F_Q^{\t{(ii/iii)}}  \lesssim \frac{  N  {\eta}}{1 -{\eta}}.
\end{equation}
Importantly, this bound is
asymptotically saturable for both models with a scheme (ii) where the
optimal input probes are prepared in a spin-squeezed states for atomic
systems \cite{kitagawa, luiz}, or in squeezed states of light for
optical implementations \cite{guta, demkowicz2013, demkowicz2014}.

In order to inspect the benefits of entangled-based strategies over
sequential ones we plot in Fig.~\ref{f:comparison} the ratio of
formulas in Eqs.~\eqref{eq:entbasedbound} and \eqref{eq:seqprec} as a
function of ${\eta}$. Note that the entanglement-enhancement factor is bounded by bounded by $\exp(1)$, a result
known in frequency estimation schemes in the limit vanishing interrogation times \cite{chiara, luiz}, which in
our case corresponds to ${\eta} \rightarrow 1$. We stress however, that in the noiseless case ${\eta}=1$ all four
metrology schemes perform equally well, achieving the Heisenberg scaling.
Finally, regarding scheme (iv), we note that since the asymptotic
bound on $F_Q^{\t{(iv)}}$ coincides with the bound on
$F_Q^{\t{(ii/iii)}}$ (as $\beta=0$) and the latter is
asymptotically saturable using (ii) for erasure and dephasing, this
immediately implies that there is no asymptotic benefit in using (iv)
in these cases.
\begin{figure}[ht]
\begin{center}
\epsfxsize=.8\hsize\leavevmode\epsffile{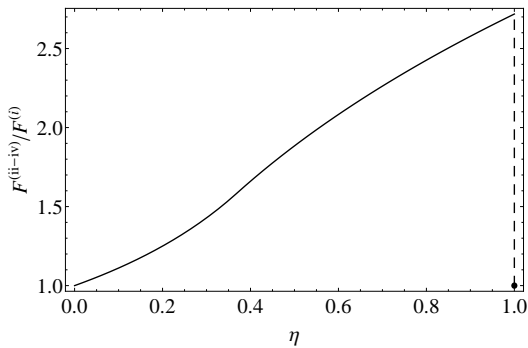}
\end{center}
\vspace{-.5cm}
\caption{Advantage of entangled-based over entanglement-free schemes
  for erasure, dephasing and amplitude damping, quantified as an asymptotic ratio of
  achievable quantum QFIs as a function of the
  decoherence parameter ${\eta}$. For ${\eta} \rightarrow 1$ the ratio
  approaches $\exp(1)$, but for the perfectly noiseless case $\eta=1$, the advantage
  vanishes which is depicted by a dot. In case of amplitude damping a further improvement is possible (bounded by a factor of 4) when using (iii,iv)
  strategies instead of (ii).}
\labell{f:comparison}\end{figure}


One can also derive the corresponding bound for the amplitude damping (see supplemental material)
which reads $F_Q^{\t{(ii/iii)}}  \lesssim \frac{4 N  {\eta}}{1 -{\eta}}$. This bound, however, is not tight
for (ii) strategies, which has been proven recently in \cite{knysh2014} using an alternative method based on the calculus of
variations---the actual tight bound for (ii) in fact coincides with Eq.~\eqref{eq:entbasedbound}. This makes the case of amplitude damping
distinct from the other two and opens up a possibility of proving the asymptotic benefits of using the ancillas, see below.

\begin{figure}[ht]
\begin{center}
\epsfxsize=.8\hsize\leavevmode\epsffile{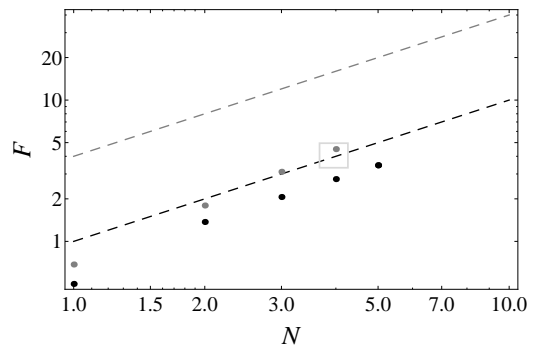}
\end{center}
\vspace{-.5cm}
\caption{Comparison between the yield of the amplitude-damping channel
  with and without passive ancillas for exemplary decoherence
  parameter $\eta=0.5$ as function of the number $N$ of maps employed in the estimation:
  attainable QFI without ancillas, strategy (ii) (black circles);
  attainable QFI with passive ancillas, strategy (iii)  (gray circles);
  asymptotically tight upper bound for the QFI for (ii) strategies from \cite{knysh2014} (dashed black line);
  our universal bound for QFI for both passive (iii) and active
  (iv) ancillas (dashed gray curve), no strategy can achieve better precision. The gray
  box emphasizes that for $N=4$, and hence also asymptotically, the
  strategy (iii) can beat the bound for all strategies of
  type (ii). (More details in the supplemental material.)}
\labell{f:amplitude}\end{figure}

Analyzing the role of ancillas, we have already shown that
they are useless in case of dephasing and erasure. Surprisingly, this not so in the case
of the amplitude-damping noise.
In this case, as mentioned above and proven in \cite{knysh2014}, bound
\eqref{eq:entbasedbound} is tight for (ii). A numerical search for
optimal ancilla assisted strategies (iii) for small number of probes
$N \leq 4$ gives a QFI that exceeds the bound \eqref{eq:entbasedbound}
for $\eta \lesssim 0.5$, see Fig.~\ref{f:amplitude}.  Most
importantly, this advantage of (iii) over (ii) strategies will be
preserved also in the asymptotic limit, since the bound
\eqref{eq:entbasedbound} is linear in $N$ and the same linear gain can
be achieved by simply repeating experiment, e.g. using the optimal $4$
particle strategy $N/4$ times.  This gives a (numerical) proof that
(ii) is strictly less powerful than (iii) for amplitude-damping.

In conclusion, we have presented a hierarchy for the performance of
quantum metrology in the presence of dephasing, erasure and
amplitude-damping noise, and illustrated a conjecture on how this
hierarchy can be extended to arbitrary noise models, based on new
general bounds. In this hierarchy, entanglement-free schemes perform
worse than entangled ones, and in some cases schemes with passive
ancillas perform better than unaided ones, even though they are all
equivalent in the noiseless case.

RDD thanks Marcin Jarzyna for useful discussions. This research work
supported by the FP7 IP project SIQS co-financed by the Polish
Ministry of Science and Higher Education.

\appendix
\section{Supplemental Material}

\subsection{Optimal single particle probe states for the dephasing noise model}
Here we prove that the $\ket{+}$ input probe state is indeed optimal for the dephasing channel
defined by Kraus representations \eqref{cpmap}.
This input probe yields  $F[\Lambda_\varphi(\ket{+}\bra{+})]=\eta$.
To see that this is indeed
the optimal probe, it is enough to show that it achieves the bound
\eqref{eq:minkraus} for a particular Kraus decomposition of the
channel. The canonical Kraus representations $K_i^\varphi =
K_i U_\varphi$, with $K_i$ given in \eqref{cpmap} is not
appropriate as the corresponding $4 \| \sum_k \dot{K}^{\varphi
  \dagger}_k \dot{K}^{\varphi}_k \|$ equals $1$.  If we, however, replace $K_i$
with $\tilde{K}_i$ given by
\begin{align}
\label{eq:optkraussingle}
\tilde{K}_0 &= \cos(\xi \varphi) K_0
- i \sin(\xi \varphi) K_1, \\
\label{eq:optkraussingle1}
\tilde{K}_1 &= \cos(\xi \varphi) K_1 - i \sin(\xi \varphi)K_1,
\end{align}
with $\xi =\sqrt{1-\eta}$, we get $4 \| \sum_k
\dot{\tilde{K}}^{\varphi \dagger}_k \dot{\tilde{K}}^{\varphi}_k
\|=\eta$ proving that $\ket{+}$ is indeed optimal.

\subsection{Optimal Kraus representations yielding asymptotically tight bounds for dephasing and erasure noise models}
\label{sec:optimalkraus}
Here we present explicit Kraus representations for the dephasing,
and erasure  noise that yield asymptotically tight bounds on QFI
given in Eq.~\eqref{eq:entbasedbound} of the main text. Detailed discussion of amplitude damping model is
given in \ref{sec:ampdamp}. All Kraus
operators are assumed to be additionally multiplied on the r.h.s. by
the unitary evolution $U_\varphi = \ket{0}\bra{0}+ e^{i
  \varphi}\ket{1}\bra{1}$, i.e.: $K^\varphi_k = K_k u_\varphi$ before
being used to calculate the bound:
 \begin{equation}
 \label{eq:asymptotic}
 F^{\t{(ii-iv)}} 4 \min_{K_k^\varphi, \beta=0} N \|\alpha\|,
 \end{equation}
 where
\begin{equation}
\alpha = \sum_k \dot{K}^{\varphi \dagger}_k \dot{K}^{\varphi}_k, \quad \beta= \sum_{k} \dot{K}^{\varphi \dagger}_k {K}^{\varphi}_k.
\end{equation}

\subsubsection{Dephasing}
While the canonical Kraus representations is given by \eqref{cpmap},
the optimal Kraus representation that gives minimal $\|\alpha\|$ under
the constraint $\beta=0$ reads
\begin{eqnarray}
\tilde{K}_0  &= \cos(\chi \varphi)
 K_0
 - i \sin(\chi \varphi) K_1 \\
 \tilde{K}_1  &=  \cos(\chi \varphi) K_1-
 i \sin(\chi \varphi)K_0
\end{eqnarray}
where $\chi=1/[2\sqrt{1-\eta}]$ and yields
$\|\alpha\|=\eta/[4(1-\eta)]$ which reproduces the asymptotically
tight bound given in Eq.~\eqref{eq:entbasedbound}. Note the
difference between this representation and the one used to derive the
optimal single probe QFI given in Eqs.~\eqref{eq:optkraussingle} and
\eqref{eq:optkraussingle1} amount to a different ``Kraus rotation
speed'': $\chi$ instead of $\xi$, which guarantees that $\beta=0$.

\subsubsection{Erasure}
The canonical Kraus operators for the erasure map is
\begin{eqnarray}
&K_0 = \mat{ccc}{\sqrt{\eta} & 0 & 0 \\ 0 & \sqrt{\eta} & 0 \\ 0 & 0 & 0},
&K_1 = \mat{ccc}{0 & 0 & 0 \\ 0 & 0& 0 \\ 0 & 0 & 1} \nonumber\\
&K_2 = \mat{ccc}{0 & 0 & 0 \\ 0 & 0 & 0 \\ \sqrt{1-\eta} & 0 & 0},
&K_3 = \mat{ccc}{0 & 0 & 0 \\ 0 & 0& 0 \\ 0 & \sqrt{1-\eta} & 0}
\end{eqnarray}
where the third dimension corresponds to the state from the phase
insensitive subspace, $K_0, K_1$ correspond to the probe remaining
untouched within the phase-sensitive or phase-insensitive subspace
respectively while $K_2, K_3$ represent the events of probe being
erased to the phase insensitive state from either $\ket{0}$ or
$\ket{1}$.  The optimal Kraus representation reads:
\begin{equation}
\tilde{K}_0  = K_0, \ \tilde{K}_1 = K_1, \tilde{K}_2 = e^{-i \zeta \varphi} K_2,\tilde{K}_3 = e^{i \zeta \varphi} K_3
\end{equation}
where $\zeta= 1/[2(1-\eta)]$ and yields $\|\alpha\|=\eta/[4(1-\eta)]$
in agreement with Eq.~\eqref{eq:entbasedbound}.


\subsection{Asymptotic bounds and inequivalence of (ii) and (iii) for amplitude-damping noise model}
\label{sec:ampdamp}
Here we discuss the asymptotic bounds for amplitude-damping noise model and show numerically that (ii) and (iii) are inequivalent
as there is a finite gap in precision
between (ii) and (iii). Making use of the formula \eqref{eq:boundsii} we obtain the asymptotic bound in the form \cite{guta,kolodynski2013}
\begin{equation}
\label{eq:ampboundiii}
F^{\t{(ii-iv)}} \leq \frac{4 N \eta}{1-\eta},
\end{equation}
which corresponds to the following optimal choice of Kraus operators $\tilde{K}_i$ expressed in terms of canonical Kraus operators
given in Eq.~\eqref{eq:krausdamping}:
\begin{equation}
\tilde{K}_1 = e^{- \mathrm{i} \varphi/ 2}K_1, \ \tilde{K}_2 = e^{\mathrm{i} \xi \varphi/2} K_2,
\end{equation}
where $\xi =  (1+\eta)/(1-\eta)$, which yields $\| \alpha \| = \eta/(1-\eta)$. Using alternative method methods of
variational calculus a tighter and asymptotically saturable bound for (ii) strategies
has been derived in \cite{knysh2014}:
\begin{equation}
\label{eq:ampboundii}
F^{\t{(ii)}} \lesssim  \frac{N \eta}{1 -\eta},
\end{equation}
which coincides with the asymptotic bounds for the dephasing and erasure noise, see Eq.~\eqref{eq:entbasedbound}.
Clearly, there is a significant gap between asymptotically tight bound \eqref{eq:ampboundii} valid for (ii) strategies and
the more general bound \eqref{eq:ampboundiii} covering all ancilla-assisted strategies.  However, since the bound
\eqref{eq:ampboundiii} is not guaranteed to be saturable even with (iii) or (iv) strategies, it is not yet a proof that there is in fact
asymptotic ancilla-assisted precision enhancement. Therefore we have performed a numerical search for the optimal (iii) strategies
for low $N$ for which the numerical search is feasible. We have achieved it by implementing a semi-definite program
minimizing the right-hand side of \eqref{eq:minkraus} over Kraus representations in a analogous way as in \cite{guta, kolodynski2013}, but
this time without assuming the product Kraus representation structure $K^{\varphi (\t{ii/iii})}_{\boldsymbol{k}} =
K^\varphi_{k_N}\otimes \dots \cdot K^\varphi_{k_1}$ for $N$ parallel channels $\Lambda_\varphi^{\otimes N}$,
but rather allowing for an arbitrary Kraus representation. This complicates numerics significantly and is therefore feasible
only for small $N$, but guarantees that the resulting value of QFI is achievable using passive-ancilla assisted strategies (iii).

In Fig.~\ref{f:amplitude} of the main text we depict the results for $\eta=0.5$, where it is evident that numerically obtained QFI for the (iii) schemes (gray circles) surpasses the bound for (ii)  (dashed, black) strategies already for $N=4$.
As argued in the main text, although this advantage is demonstrated here only for finite $N=4$, it can be
pushed to the interesting asymptotic regime of infinite $N$ by simply
repeating the experiment many times independently and averaging the
outcomes. Note also, that by merely analyzing $N=1$ case for which the analytical formula for the ancilla assisted scheme
is known \cite{kolodynski2013}, and reads: $F=4\eta/(1+\sqrt{\eta})^2$, one can identify the regime of $\eta$ for which
the violation of the (ii) bound occurs already for $N=1$, as inequality $4\eta/(1+\sqrt{\eta})^2  >  \eta/(1-\eta) $
holds for $\eta < 0.36$.  This proves that indeed (ii) is
inequivalent to (iii): surprisingly the use of passive ancillas allows
one to achieve a strictly higher accuracy in estimation with
amplitude-damping noise. For comparison we also present numerically obtained results
of achievable QFI in case of (ii) strategies (black circles) which were obtained using a numerical iterative algorithm proposed in
\cite{macieszczak2013, macieszczak2013b}, as well as the universally valid bound \eqref{eq:ampboundiii} (dashed, gray). It still remains an interesting
open question whether the bound \eqref{eq:ampboundiii} is tight, i.e. whether the solid gray line will asymptotically approach it,
and if not whether allowing for active adaptive schemes (iv) would be sufficient to actually reach the bound. We also cannot exclude the possibility
that the bound is simply not tight and no strategies, even the most general (iv), can approach it asymptotically.

\subsection{Derivation of the bound for general feedback assisted schemes $\t{(iv)}$}
\label{sec:derivation}
 Let us first try to derive a bound for a simple sequential strategy (i) ($k=1, n=N$) making use of Eq.~\eqref{eq:minkraus} and a
  product Kraus representation $K^{\varphi (\t{i})}_{\boldsymbol{k}} = K^\varphi_{k_N}\cdot \dots \cdot K^\varphi_{k_1}$:
 \begin{multline}
 F^{(\t{i})} \leq 4 \min_{K_k^\varphi} \| \sum_{\boldsymbol{k}} \dot{K}^{\varphi (\t{i})\dagger }_{\boldsymbol{k}}
 \dot{K}^{\varphi (\t{i})}_{\boldsymbol{k}} \| = \\
 4 \min_{K_k^\varphi}\| \sum_{\boldsymbol{k}} \sum_{i,j=1}^N K^{\varphi \dagger}_{k_1}  \dots
  \dot{K}^{\varphi \dagger}_{k_i}\dots K^{\varphi \dagger}_{k_N}  K^{\varphi }_{k_N}  \dots \dot{K}^{\varphi \dagger}_{k_j}
 \dots  K^{\varphi}_{k_1} \|.
 \end{multline}
 First note  the following property of the operator norm
 \begin{equation}
 \label{eq:normineq}
 \|{\sum_k L^\dagger_k A L_k}\| \leq \| A \|\,  \| \sum_k L^\dagger_k L_k \|
 \end{equation}
 valid for any operator $A$ and any set of
 operators $L_k$. Making use of the above monotonicity property together with the triangle inequality and
  the trace preservation condition $\sum_k K_k^{\varphi \dagger} K_k^{\varphi} = \openone$ we get:
\begin{multline}
\label{eq:fishbound1}
F^{(\t{i})} \leq 4 \min_{K_k^\varphi} \sum_i \| \sum_{k_i}\dot{K}^{\varphi \dagger}_{k_i} \dot{K}^{\varphi \dagger}_{k_i} \| + \\
  +\sum_{i<j} \| \sum_{k_i,\dots,k_j}\dot{K}^{\varphi \dagger}_{k_i}\dots {K}^{\varphi \dagger}_{k_j} \dot{K}^{\varphi}_{k_j}
  \dots {K}^{\varphi}_{k_i}  + h.c.\|.
\end{multline}
Focusing on the second summation term, observe that the trace preservation condition implies that
$\sum_k \dot{K}_k^\dagger K_k$ is anti-hermitian. Let $i A $ be an anti-hermitian operator and consider the following
chain of inequalities
\begin{multline}
\|\sum_k \dot{K}^\dagger_k  i A K_k + h.c.\|  = \| i \sum_k \dot{K}^\dagger_k  A K_k  -  {K}^\dagger_k  A \dot{K}_k  \|  = \\
= \| \sum_k (\dot{K}_k + i K_k)^\dagger A (\dot{K}_k + i K_k) - \dot{K}^\dagger_k  A \dot{K}_k - {K}^\dagger_k  A K_k \|.
\end{multline}
Making use of the triangle inequality together with Eq.~\eqref{eq:normineq} we upper bound the above expression by
\begin{multline}
 \leq 2 \| A\|\left (\|\sum_k \dot{K}^{\dagger}_k \dot{K}_k\| + \| \sum_k \dot{K}^\dagger_k K^\dagger_k \| +1\right).
\end{multline}
The above inequality allows us to rewrite Eq.~\eqref{eq:fishbound1} in
a final form:
\begin{equation}
\label{eq:unversalbound}
F^{\t{(i)}} \leq 4 \min_{K_k^\varphi} N \|\alpha\| + N(N-1) \|\beta\| (\|\alpha\| + \| \beta\| + 1),
\end{equation}
where
\begin{equation}
\alpha = \sum_k \dot{K}^{\varphi \dagger}_k \dot{K}^{\varphi}_k, \quad \beta= \sum_{k} \dot{K}^{\varphi \dagger}_k {K}^{\varphi}_k.
\end{equation}
All the steps in the above derivation are also valid for (iv) in
addition to (i), as the intermediate unitary operations $U_i$ do not
affect the values of the operator norms that appear in the
derivation. Hence, inequality \eqref{eq:unversalbound}  applies also
to this case:
\begin{equation}
\label{eq:unversalboundiv}
F^{\t{(iv)}} \leq 4 \min_{K_k^\varphi} N \|\alpha\| + N(N-1) \|\beta\|
(\|\alpha\| + \| \beta\| + 1).
\end{equation}

This should sound an alarm that most probably the bound is far from
tight for (i), as it is invariant under such a significant
generalization of the scheme. This is indeed the case.  As discussed
in the main text, for the case of erasure and dephasing decoherence
models the above bound is asymptotically tight when (iv) schemes are
considered, but is hardly useful in analyzing the performance of (i).
The intuitive reason behind this, is that when minimizing over Kraus
representations of the channels we have restricted ourselves to
\emph{product} Kraus representations derived from single channel Kraus
representations. When thinking of $\Lambda^N_\varphi$ this makes an
artificial formal separation of the channels acting in a sequence, and
significantly better bounds may be derived by performing a
minimization over general Kraus representation of $\Lambda^N_\varphi$
instead of just product ones.

\end{document}